\documentclass[reprint,superscriptaddress,amsmath,amssymb,amsfonts,aps,pra]{revtex4-1}

\usepackage{amsmath}
\usepackage{hyperref}
\usepackage{bm}
\usepackage{dsfont}
\usepackage{amsthm}
\usepackage{amsmath}
\usepackage{verbatim} 
\usepackage{qcircuit}
\usepackage{circuitikz}
\usepackage{tabularx}
\usepackage{color,soul}
\usepackage{esint}
\usepackage{braket}
\usepackage{bbold}
\usepackage{amssymb}
\usepackage{float}

\begin{document}

\author{Andrey Kardashin}
\email{andrey.kardashin@skoltech.ru}\homepage{http://quantum.skoltech.ru}
\affiliation{Skolkovo Institute of Science and Technology, Moscow 121205, Russia}
\author{Anastasiia Pervishko}
\affiliation{Skolkovo Institute of Science and Technology, Moscow 121205, Russia}
\author{Jacob Biamonte}
\affiliation{Skolkovo Institute of Science and Technology, Moscow 121205, Russia}
\author{Dmitry Yudin}
\affiliation{Skolkovo Institute of Science and Technology, Moscow 121205, Russia}

\title{Numerical hardware-efficient variational quantum simulation of a soliton solution}

\begin{abstract}
    Implementing variational quantum algorithms with noisy intermediate-scale quantum machines of up to a hundred of qubits is nowadays considered as one of the most promising routes towards achieving a quantum practical advantage. In multiqubit circuits, running advanced quantum algorithms is hampered by the noise inherent to quantum gates which distances us from the idea of universal quantum computing. Basing on a one-dimensional quantum spin chain with competing symmetric and asymmetric pairwise exchange interactions, herein we discuss the capabilities of quantum algorithms with special attention paid to a hardware-efficient variational eigensolver. A delicate interplay between magnetic interactions allows one to stabilize a chiral state that destroys homogeneity of magnetic ordering, thus making this solution highly entangled. Quantifying entanglement in terms of quantum concurrence, we argue that, while being capable of correctly reproducing a uniform magnetic configuration, the hardware-efficient ansatz meets difficulties in providing a detailed description to a noncollinear magnetic structure. The latter naturally limits the application range of variational quantum computing to solve quantum simulation tasks.
\end{abstract}

\maketitle

{\it Introduction.} Combining different aspects of algorithm development with quantum engineering is regarded nowadays as a feasible tool to accelerate computations \cite{McClean2016,Babbush2016,Yang2017,Paesani2017,Li2017,Dunjko2018,Preskill2018,Hempel2018,Colless2018,Santagati2018,Babbush2018,Kivlichan2018,Moll2018,LaRose2019,Schuld2019,Huggins2019,Gyongyosi2019,Cross2019,McArdle2019,Lee2019,Yuan2019,Zhu2019,Wang2019,Liu2019,Carolan2020,Uvarov2020,McArdle2020,Schuld2020,Endo2020,Lubasch2020,Kardashin2020,Cerezo2021,Biamonte2021,Monroe2021,Alexeev2021,Harrow2021,Skolik2021}. One of the most promising classes of algorithms for noisy intermediate-scale quantum (NISQ) devices of up to a hundred of qubits are the hybrid quantum-classical algorithms \cite{peruzzo_variational_2014, Kokail2019, yung2014transistor} that enjoy a classical outer loop optimizer, where a measured objective function is minimized iteratively, in terms of structure and depth of the ansatz state as well as penalty function. This approach is based on distributing the computational routines between a classical and quantum computer, taking into account that some of these routines can be executed on one kind of device more efficiently than on the other. A typical example is the variational quantum eigensolver (VQE) \cite{peruzzo_variational_2014}. Given an $n$-qubit Hamiltonian $H$, this algorithm allows one to find its lowest-lying eigenvalue and the corresponding eigenvector. In VQE, one uses a quantum computer for preparing a probe state $\ket{\psi(\boldsymbol{\theta})}$, which is parametrized by a set of $p$ angles $\boldsymbol{\theta} \in [0, 2\pi)^{\times p}$, and measures the expectation value of the given Hamiltonian in this state, $\braket{\psi(\boldsymbol{\theta}) \vert H \vert \psi(\boldsymbol{\theta})}$. A classical computer, in its turn, is used to update the parameters $\boldsymbol{\theta}$ by means of some optimization method in order to minimize the expectation value. The variational state is usually prepared by acting with a parametrized unitary operator $U(\boldsymbol{\theta})$ on the initial state $\ket{0}^{\otimes n}$ or any other easy-to-prepare state, so that $\ket{\psi(\boldsymbol{\theta})} = U(\boldsymbol{\theta}) \ket{0}^{\otimes n}$. The unitary $U(\boldsymbol{\theta})$ is essentially a quantum circuit specified by a chosen ansatz; in practice, unitary coupled cluster \cite{taube2006new,Shen2017}, tensor networks state \cite{kardashin2021quantum,Huggins2019}, and hardware-efficient ansatz  \cite{kandala_hardware-efficient_2017} are among the most popular options.

Quantum entanglement that describes nonclassical correlations between spatially separated parts of a system endows a quantum computer with the advantage to execute multiple computation tasks in parallel. In this respect, studying entanglement in quantum spin chains provides us with a unique tool to test contemporary quantum algorithms. In practice, one can address the relationship between families of variational quantum circuit ans\"atze and families of objective functions (Hamiltonians) these circuits can minimize. The two most studied quantum spin models are the transverse field Ising model \cite{Schultz1964,Pfeuty1970,Dutta2015,Arai2018,Berezutskii2020} and anisotropic Heisenberg chain \cite{Baxter1972}. In the meantime, recently it was demonstrated that the Dzyaloshinskii-Moriya interaction (DMI) drastically modifies the behavior of entanglement in a one-dimensional quantum spin chain \cite{Kargarian2009,Radhakrishnan2017,Soltani2019,Yi2019,Thakur2020,jafari2011three,mehran2014induced,jafari2008phase}. Indeed, DMI, derived first by Dzyaloshinskii on purely phenomenological grounds \cite{Dzyaloshinsky1958}, serves as a source of magnetic frustration resulting in neighboring magnetic moments being arranged in a spiral, thus making the ground state more entangled as opposed to collinearly ordered. In the following, it was pointed out by Moriya that DMI might be derived in a perturbative manner from the Anderson's superexchange theory provided spin-orbit coupling is included \cite{Moriya1960}.

\begin{figure*}
    \includegraphics[width=.75\textwidth]{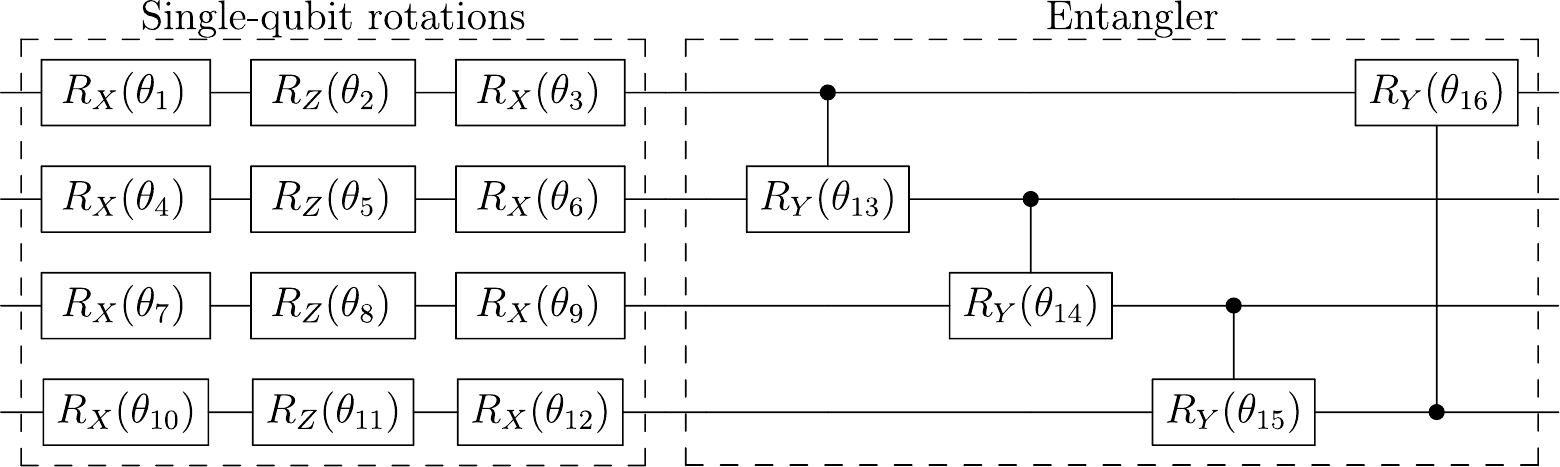}
    \caption{A quantum circuit that represents a layer of the hardware-efficient ansatz (16 variational parameters for a set of four qubits) used throughout our numerical simulations. The single-qubit rotations block is constituted by a sequence of $X$, $Z$, and $X$ rotations, whereas the entangling block is equipped with controlled $Y$ rotations. Note that $R_\alpha(\theta) = e^{-\imath \theta\hat{\sigma}_\alpha}$ ($\alpha \in \{X,Y,Z\}$) with $\hat{\sigma}_\alpha$ being the corresponding Pauli matrix, and $\theta_j \in [0, 2\pi)$. To increase the expressive power of the ansatz, more layers can be added.}
    \label{fig:hardware_efficient_ansatz}
\end{figure*}

In this Letter, we employ the numerical hardware-efficient VQE to analyze ground state properties of a ferromagnetic Heisenberg chain with DMI in a transverse magnetic field. Practically, we demonstrate that VQE underperforms when approximating a noncollinear magnetic structure. To provide a quantitative estimate we analyze entanglement properties of the VQE solution by a means of quantum concurrence, that is purely determined by a two-qubit reduced density matrix \cite{Hill1997,Wootters1998,Horodecki2009}. The last but not least, by using the VQE solution we show how the spin configuration evolves with increasing the number of layers in the ansatz state. An interesting observation is that a one-layer VQE solution reproduces the spin configuration that agrees well with an exact analytical solution as obtained in the continuum limit. 

{\it Model system.} Consider the Hamiltonian of a one-dimensional chain of $N$ interacting quantum spins $\hat{\mathbf{S}}_j$ labeled by their position $j$ along the $z$ axis,
\begin{equation}\label{eq:inham}
    \hat{H}=-J\sum\limits_{\langle i,j\rangle}\hat{\mathbf{S}}_i\cdot\hat{\mathbf{S}}_j-\sum\limits_{\langle i,j\rangle}\bm{D}_{ij}\cdot(\hat{\mathbf{S}}_i\times\hat{\mathbf{S}}_j)+\sum\limits_{j=1}^N\mathbf{B}\cdot\hat{\mathbf{S}}_j,
\end{equation}
where the first term describes direct exchange interaction which for $J>0$ favors ferromagnetic ordering. Present in magnetic structures with a lack of inversion symmetry, DMI, specified by the second term, destroys the homogeneity of collinear magnetic ordering by promoting spin canting between neighboring sites. The Dzyaloshinskii vector $\bm{D}_{ij}$ determines the strength of DMI. The competition between Heisenberg exchange and DMI results in a noncolinear ground state configuration being stabilized in a transverse magnetic field $\mathbf{B}$, the last contribution to (\ref{eq:inham}) \cite{Togawa2012,Kishine2015,Koumpouras2016}. Note that summation over nearest neighbors $\langle i,j\rangle$ is implied and $\mathbf{B}$ is expressed in energy units. The Hamiltonian as given by (\ref{eq:inham}) provides a reliable model description to a wide class of chiral magnets, and Cr$_{1/3}$NbS$_2$ is a practical example \cite{Moriya1982,Miyadai1983,Dyadkin2015}. The hexagonal structure of this compound is composed from NbS$_2$ layers intercalated by Cr ions, thus exchange interaction and DMI emerge between Cr ions, belonging to two interacalating layers and separated by NbS$_2$. 
\begin{figure*}
    \includegraphics[width=.495\textwidth]{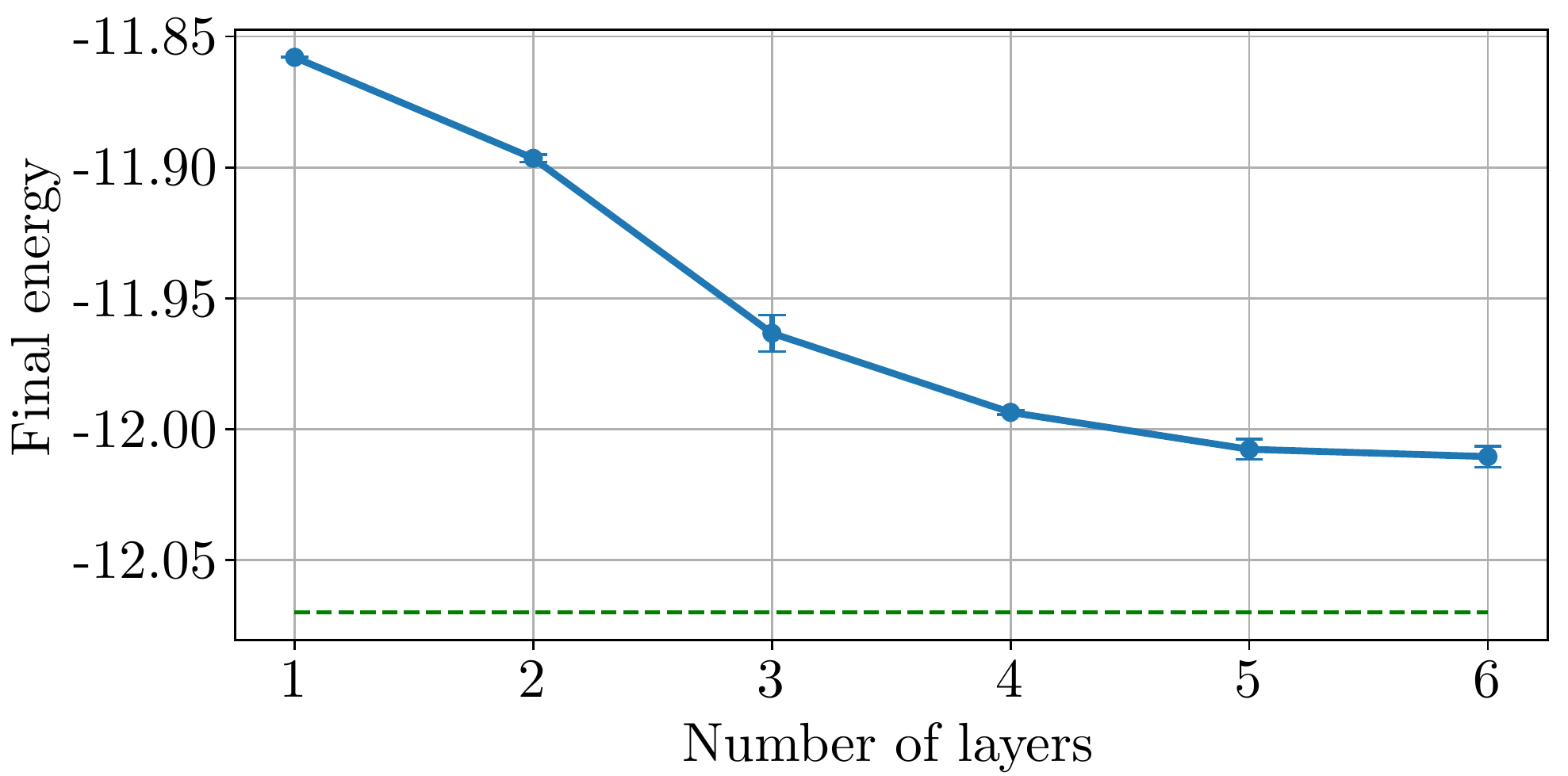}
    \includegraphics[width=.495\textwidth]{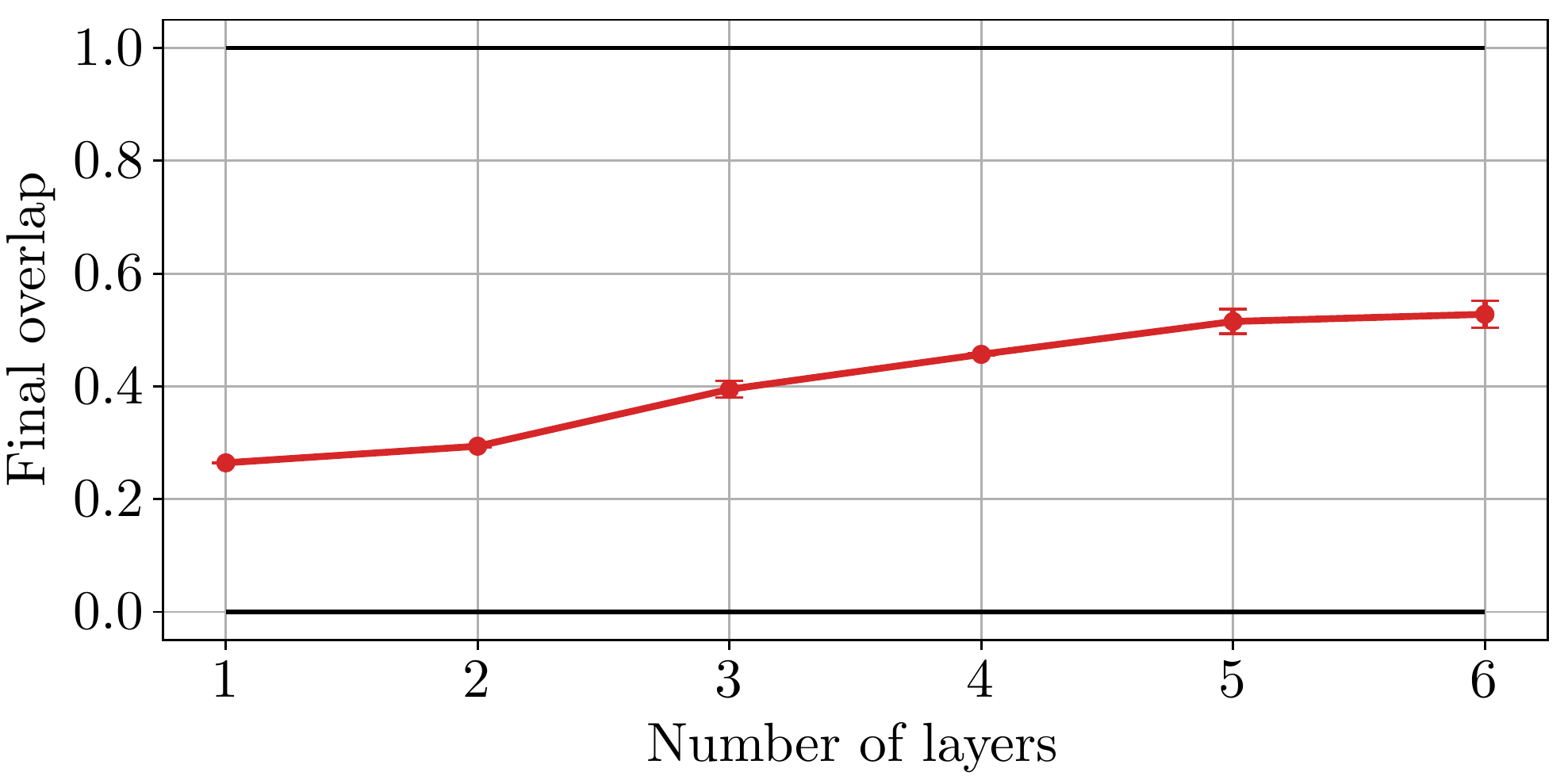}
    \caption{A numerical solution to the Hamiltonian \eqref{eq:inham} for $N=10$ qubits as implemented by means of VQE. The ground state energy and the overlap between the VQE state and the exact one depending on the number of layers in the ansatz are shown in the left and right panels, respectively. The dashed green line in the left panel marks the lowest-energy as obtained by exact diagonalization. Each data point corresponds to an average over five runs with random initial parameters of the ansatz. The plots are supplied with error bars.}
    \label{fig:energy-overlap}
\end{figure*}

For spin one-half particles, $\hat{\mathbf{S}}=\hat{\bm{\sigma}}/2$ with $\hat{\bm{\sigma}}=(\hat{\sigma}_x,\hat{\sigma}_y,\hat{\sigma}_z)$ specifying the Pauli vector. A two-component spinor $\vert S\rangle=\left(e^{-i\varphi/2}\cos\frac{\theta}{2},\; e^{i\varphi/2}\sin\frac{\theta}{2}\right)^T$, parametrized by polar $\theta$ and azimuthal angle $\varphi$, represents a quantum spin state for SU(2), so that $\langle S\vert\hat{\mathbf{S}}_i\vert S\rangle=\bm{n}_i/2$, where a unit vector $\bm{n}_i=(\cos\varphi_i\sin\theta_i,\sin\varphi_i\sin\theta_i,\cos\theta_i)$. In the following, we set the Dzyaloshinskii vector $\bm{D}_{ij}=D\hat{\bm{e}}_z$ to be aligned along the $z$ axis with the parameter $D$ determining the strength of DMI, while the magnetic field $\mathbf{B}=B\hat{\bm{e}}_x$. Thus, in the basis $\vert S_1,S_2,\ldots,S_N\rangle$ the quantum Hamiltonian (\ref{eq:inham}) can be mapped to a classical Heisenberg-type model of interacting spins, 
\begin{equation}\label{eq:clham}
    H=-\frac{J}{4}\sum\limits_{\langle i,j\rangle}\bm{n}_i\cdot\bm{n}_j-\frac{D}{4}\sum\limits_{\langle i,j\rangle}\left(\bm{n}_i\times\bm{n}_j\right)_z-\frac{B}{2}\sum\limits_{j=1}^Nn_j^x.
\end{equation}
We further proceed with a continuous description of the model (\ref{eq:clham}) in terms of magnetization specified by a unit vector field $\bm{n}(z)=[\cos\varphi(z)\sin\theta(z),\sin\varphi(z)\sin\theta(z),\cos\theta(z)]$. Note that the distance between a pair of neighboring spins $a$ determines the smallest length scale in the system, validating thus $\bm{n}(z+a)\approx\bm{n}(z)+a\bm{n}'(z)+a^2\bm{n}''(z)/2$. Replacing in (\ref{eq:clham}) the summation by integrating $\sum_j\rightarrow\frac{1}{a}\int_0^L dz$ with $L$ standing for the length of a spin chain, we derive in the lowest order in $a$:
\begin{eqnarray}\nonumber
    H=\frac{aJ}{8}\int_0^Ldz\big[\theta'^2+\varphi'^2\sin^2\theta&-&k_0\varphi'\sin^2\theta \\ \label{eq:finham}
    &+&2m^2\cos\varphi\sin\theta\big],
\end{eqnarray}
where $k_0=D/(aJ)$ is the pitch vector and $m^2=2B/(a^2J)$. The lowest-energy state of the Hamiltonian (\ref{eq:finham}) corresponds thus to $\theta=\pi/2$ on condition that $\varphi$ obeys the static sine-Gordon equation \cite{Togawa2012,Kishine2015,Koumpouras2016},
\begin{equation}\label{eq:sing}
    \varphi''+m^2\sin\varphi=0,
\end{equation}
which admits a solution in the form of a chiral soliton lattice for certain values of $D$ and $B$. From the physics point of view, the uniform magnetic field has a tendency to untwist the helical alignment of magnetic moments, that stems from a delicate interplay between the exchange interaction and DMI, towards a uniform ferromagnetic ordering via the formation of a chiral soliton lattice. Direct integration of (\ref{eq:sing}) leads to
\begin{equation}\label{eq:solution}
    \varphi=2\mathrm{am}(mz/\kappa,\kappa),
\end{equation}
where $\kappa$ is the elliptic modulus and $\mathrm{am}(u,\kappa)$ is the Jacobi amplitude that is determined by $\mathrm{sn}u=\sin\mathrm{am}(u,\kappa)$ with $\mathrm{sn}u$ defining the elliptic sine. The solution corresponds to a soliton lattice with the spatial periodicity,
\begin{equation}
    \ell=\frac{2\kappa}{m}\int_0^{\pi/2}\frac{d\varphi}{\sqrt{1-\kappa^2\sin^2\varphi}}=\frac{2\kappa}{m}K(\kappa), 
\end{equation}
where $K(\kappa)$ is the complete elliptic integral of the first kind. Plugging (\ref{eq:solution}) into the expression (\ref{eq:finham}), one derives energy of a soliton lattice over a period
\begin{equation}
    \varepsilon=\frac{am^2J}{2}\left(\frac{2}{\kappa^2}\frac{E(\kappa)}{K(\kappa)}-\frac{1}{\kappa^2}-\frac{\pi}{2m}\frac{k_0}{\kappa K(\kappa)}\right),
\end{equation}
where we introduced $E(\kappa)=\int_0^{K(\kappa)}\mathrm{dn}^2zdz$, the complete elliptic integral of the second kind. To complete the analysis, one has to identify the value of $\kappa$ that minimizes the energy $\varepsilon$:
\begin{equation}\label{eq:kap}
    \pi\kappa k_0=4mE(\kappa).
\end{equation}
Note that to deduce (\ref{eq:kap}) we made use of $\kappa E'(\kappa)=E(\kappa)-K(\kappa)$ and $\kappa K'(\kappa)=E(\kappa)/(1-\kappa^2)-K(\kappa)$. Clearly, once a soliton lattice is stabilized, Eq.~\ref{eq:kap} possesses a real-valued solution.

We proceed further with a quantum simulation of a spin chain represented by the Hamiltonian (\ref{eq:inham}) for a set of parameters that allows us to stabilize a chiral soliton lattice. In particular, we inspect whether the use of variational quantum algorithms is adequate to capture this highly entangled state. To make the results of the numerical simulations sensible, we make use of the parameters $J=1.88$~mRy, $D/J=0.63$, $B/J=3.36\times10^{-3}$, which translates to a transverse field of $0.74$~T, and the self-consistent solution to (\ref{eq:kap}) gives rise to $\kappa\approx0.256$. The latter corresponds to $N=\ell/a\approx10$, i.e., to properly address one period of a soliton lattice we have to use $N=10$ qubits.

{\it Variational quantum simulation.} Here, we present the numerical results on the lowest-energy state of the Hamiltonian \eqref{eq:inham} by virtue of VQE. To apply VQE, it necessitates to decompose the target Hamiltonian $H$ as a sum of Pauli strings, 
\begin{equation}\label{eq:hamiltonian_decomposition}
    \mathcal{H} = \sum \mathcal{J}^{ij \dots k}_{\alpha \beta \dots \gamma} \,\sigma^{i}_{\alpha} \sigma^{j}_{\beta} \dots \sigma^{k}_{\gamma},
\end{equation}
where the upper latin indices stand for a qubit's number and the lower greek indices specify a Pauli operator from $\sigma \in \{\mathbb{1}, X, Y, Z\}$. The real-valued tensor $\mathcal{J}$ specifies the multiqubit coupling strength; it is easy to see that a spin chain as given by (\ref{eq:inham}) represents a special case of the generalized model \eqref{eq:hamiltonian_decomposition}.

To parametrize our probe state, we use the hardware-efficient ansatz \cite{kandala_hardware-efficient_2017}. Essentially, this kind of ansatz contains several layers of single-qubit rotations followed by a block that entangles all qubits. In our realization, we represent single-qubit rotations as a sequence of $X$, $Z$, and $X$ rotations, while the entangling block is built up from a cascade of controlled $Y$ rotations; see Fig.~\ref{fig:hardware_efficient_ansatz} for details.

In our numerical simulations, we address the expressive power of the solution as obtained with VQE depending on the number of layers in the hardware-efficient ansatz. As explained earlier, we study the Hamiltonian \eqref{eq:inham} for $N=10$ qubits; this number was shown to capture one period of a chiral soliton lattice as long as $D/J=0.63$, $B/J=3.36\times 10^{-3}$. The numerical results are shown in Fig.~\ref{fig:energy-overlap}. A quantum circuit simulation was performed with the Qiskit package \cite{Qiskit}, while energy minimization within the VQE loop was implemented based on the Broyden-Fletcher-Goldfarb-Shanno (BFGS) algorithm \cite{NoceWrig06}. Note that for each optimization cycle the maximum number of iterations was restricted to 50 000. To quantify the precision of the VQE solution, we adopt a simple criterion discussed in \cite{BravoPrieto2020scalingof, borzenkova2021variational}. Assume $E_0$ and $E_1$ are the ground state and the first excited energies as obtained, e.g., by exact diagonalization, whereas $E^\mathrm{VQE}$ is that evaluated in VQE. For the VQE solution to be accepted one has to meet $\delta=(E^\mathrm{VQE}-E_0)/(E_1-E_0)<1$. In our simulations, $\delta\approx0.6841$. 

\begin{figure*}
    \includegraphics[width=.325\textwidth]{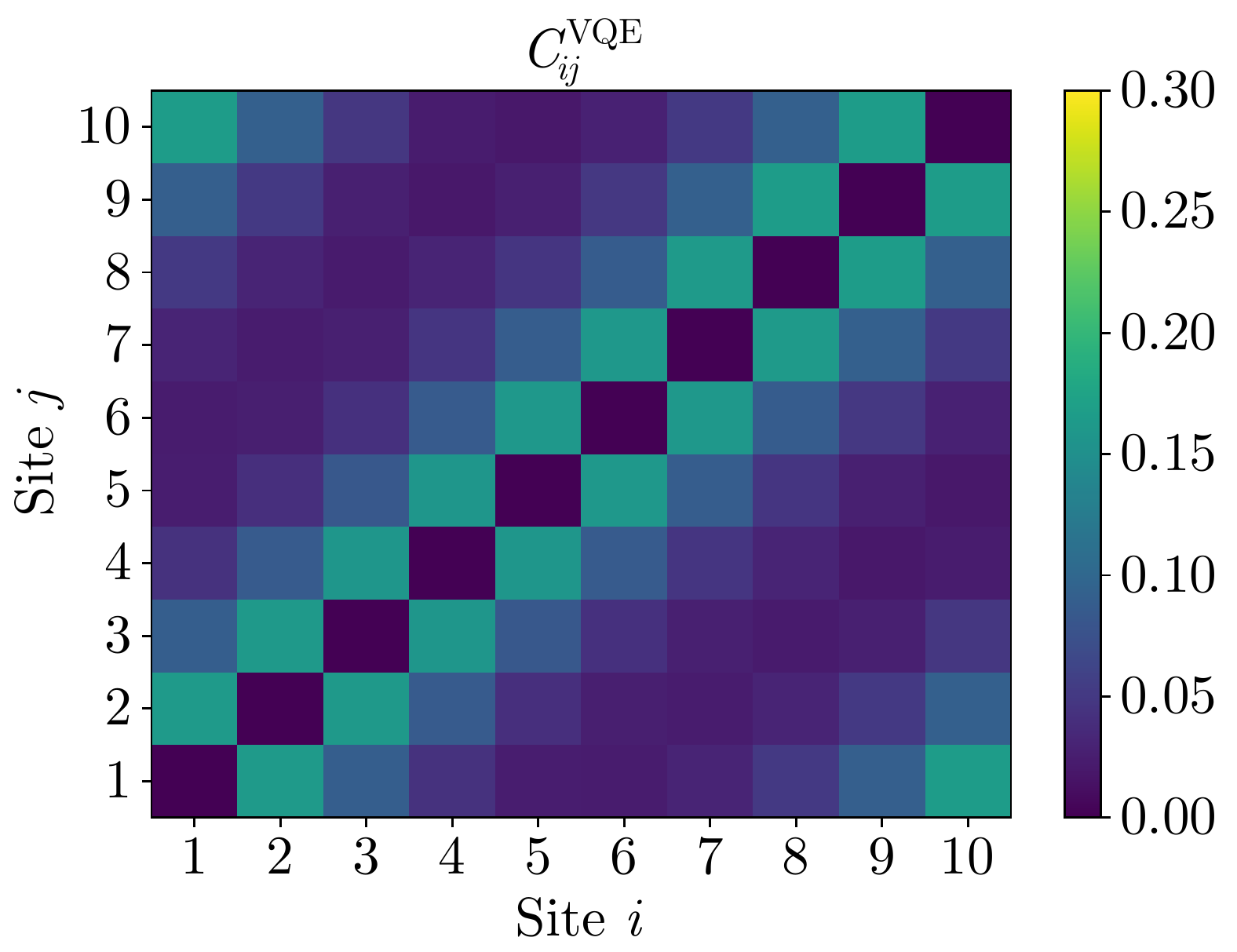}
    \includegraphics[width=.325\textwidth]{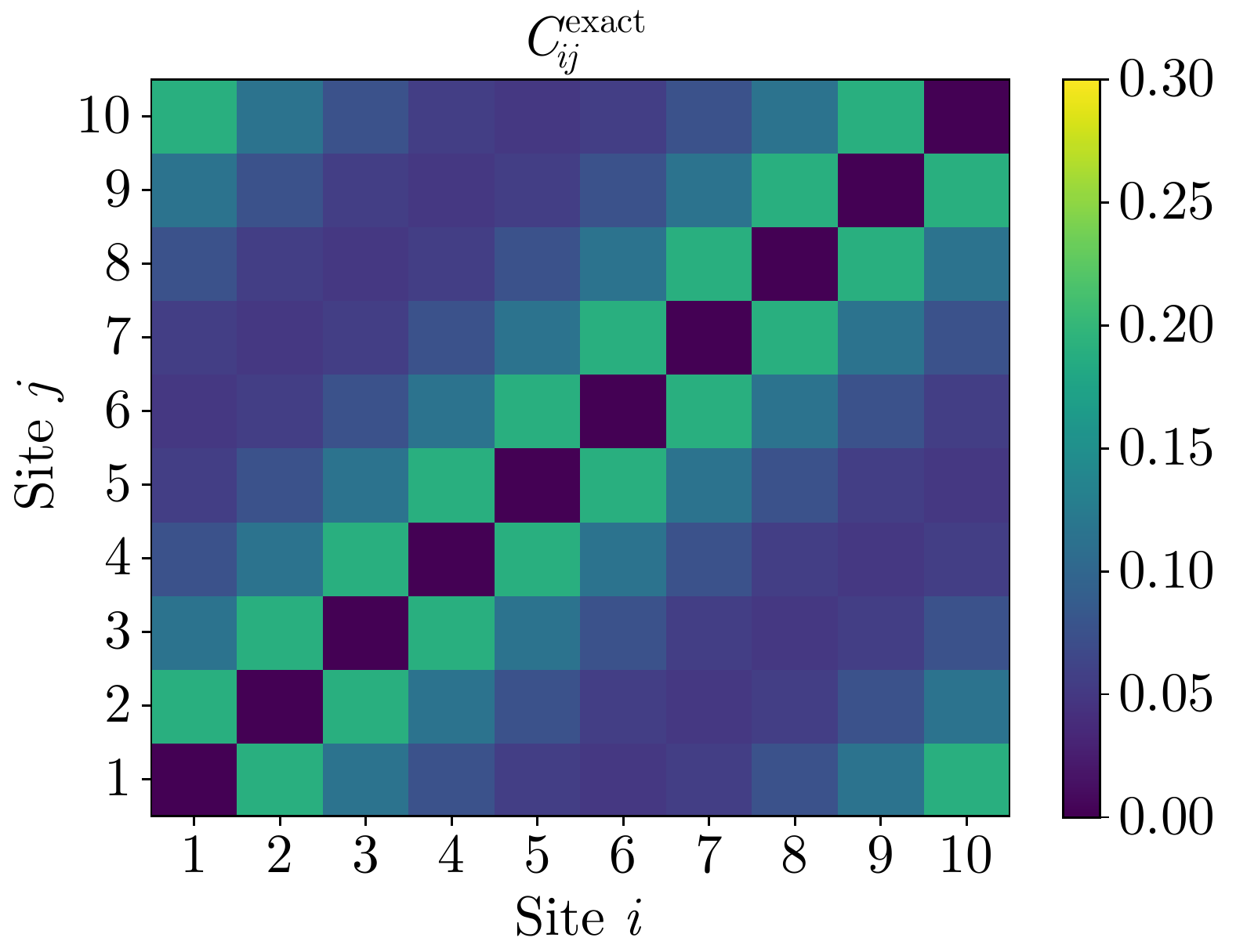}
    \includegraphics[width=.325\textwidth]{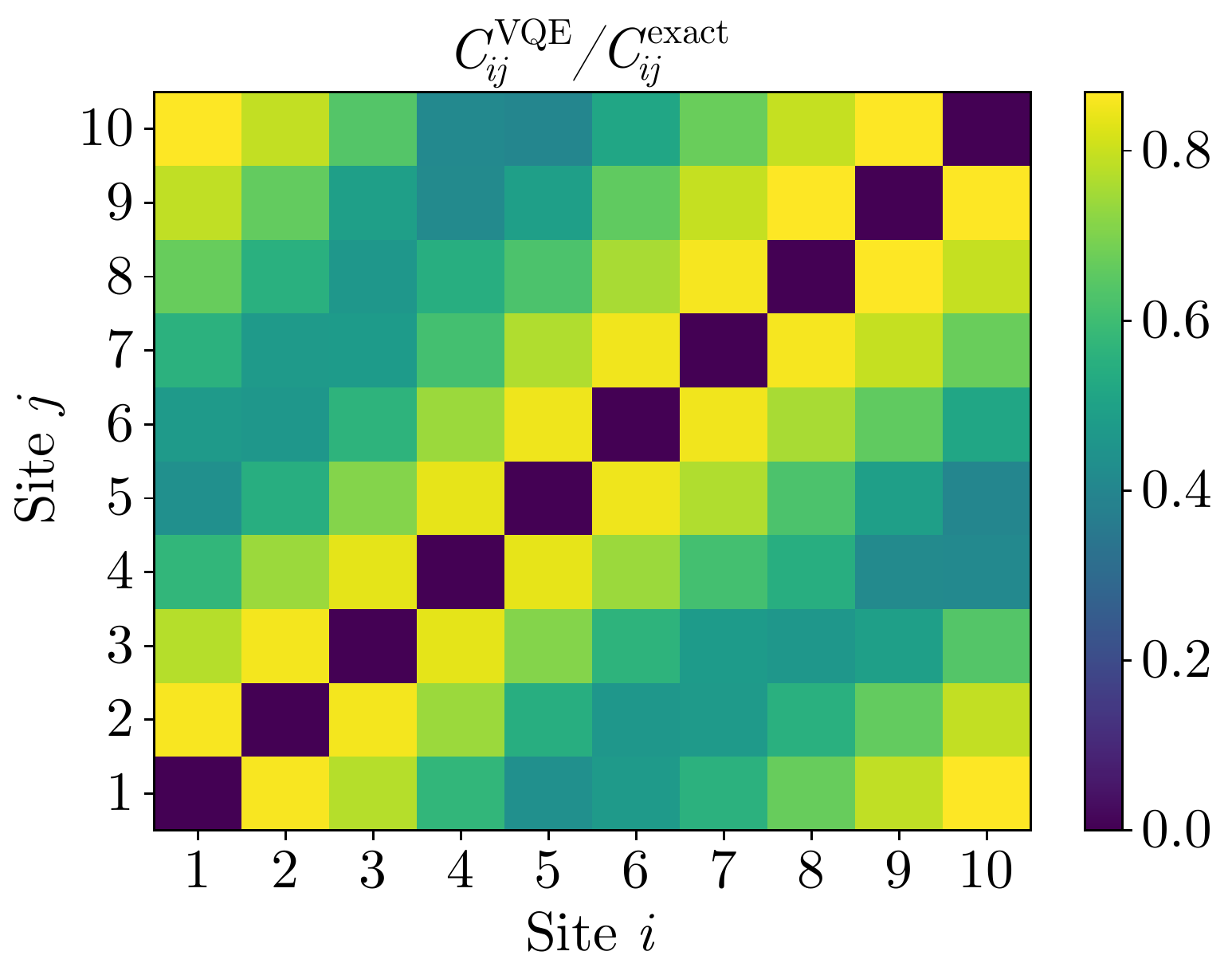}
    \caption{Concurrence between the $i$th and $j$th qubit, $C_{ij}$, estimated based on the VQE solution (left panel) and the exact solution (middle panel). Shown in the right panel is the VQE concurrence relative to the exact one. The exact concurrence $C_{ij}^\mathrm{exact}$ is evaluated based on the lowest-lying eigenstate of the Hamiltonian (\ref{eq:inham}) as obtained by exact diagonalization. Clearly, this ratio significantly varies among different regions. The VQE solution is capable of correctly keeping track of entanglement between nearest- and next-to-nearest neighboring sites, whereas the accuracy of VQE modelling dramatically decreases for sites beyond next-nearest neighbors.}
    \label{fig:concurrences}
\end{figure*}

Studying the overlap between the ground state as obtained with VQE and the exact one unambiguously reveals a poor performance of VQE when approximating a highly entangled state. In principle, entanglement properties are only determined by a many-body ground state rather than a Hamiltonian to be minimized. To provide a quantitative estimate, we adopt quantum concurrence $C_{ij}$ that measures entanglement between $i$th and $j$th sites. Given a reduced density matrix $\rho_{ij}^{(2)}$ of two qubits $i$ and $j$, one defines concurrence as
\begin{equation}\label{eq:concurrence}
    C_{ij} = \max\{0, \sqrt{\lambda_1} - \sqrt{\lambda_2} - \sqrt{\lambda_3} - \sqrt{\lambda_4}\},
\end{equation}
where $\lambda_1 \geqslant \lambda_2 \geqslant \lambda_3 \geqslant \lambda_4$ are the eigenvalues of the non-Hermitian matrix $R_{ij}=\rho_{ij}^{(2)}\tilde{\rho}_{ij}^{(2)}$ in increasing order. Here, $\tilde{\rho}_{ij}^{(2)} = (\hat{\sigma}_y \otimes \hat{\sigma}_y) \rho_{ij}^{*(2)} (\hat{\sigma}_y \otimes \hat{\sigma}_y)$ is the spin-flipped density matrix with the asterisk standing for complex conjugation. The concurrence interpolates between zero and one; two sites are completely disentangled with the rest of the system if the concurrence equals one, otherwise $i$th qubit is entangled with $j$th qubit and the other sites. In Fig.~\ref{fig:concurrences}, we provide $C_{ij}$ for the Hamiltonian \eqref{eq:inham} of $N=10$ spins based on the VQE solution. As expected, the concurrence between nearest-neighbouring spins is characterized by maximal values, meaning that these sites are the most entangled.

\begin{figure}[h!]
    \includegraphics[width=.45\textwidth]{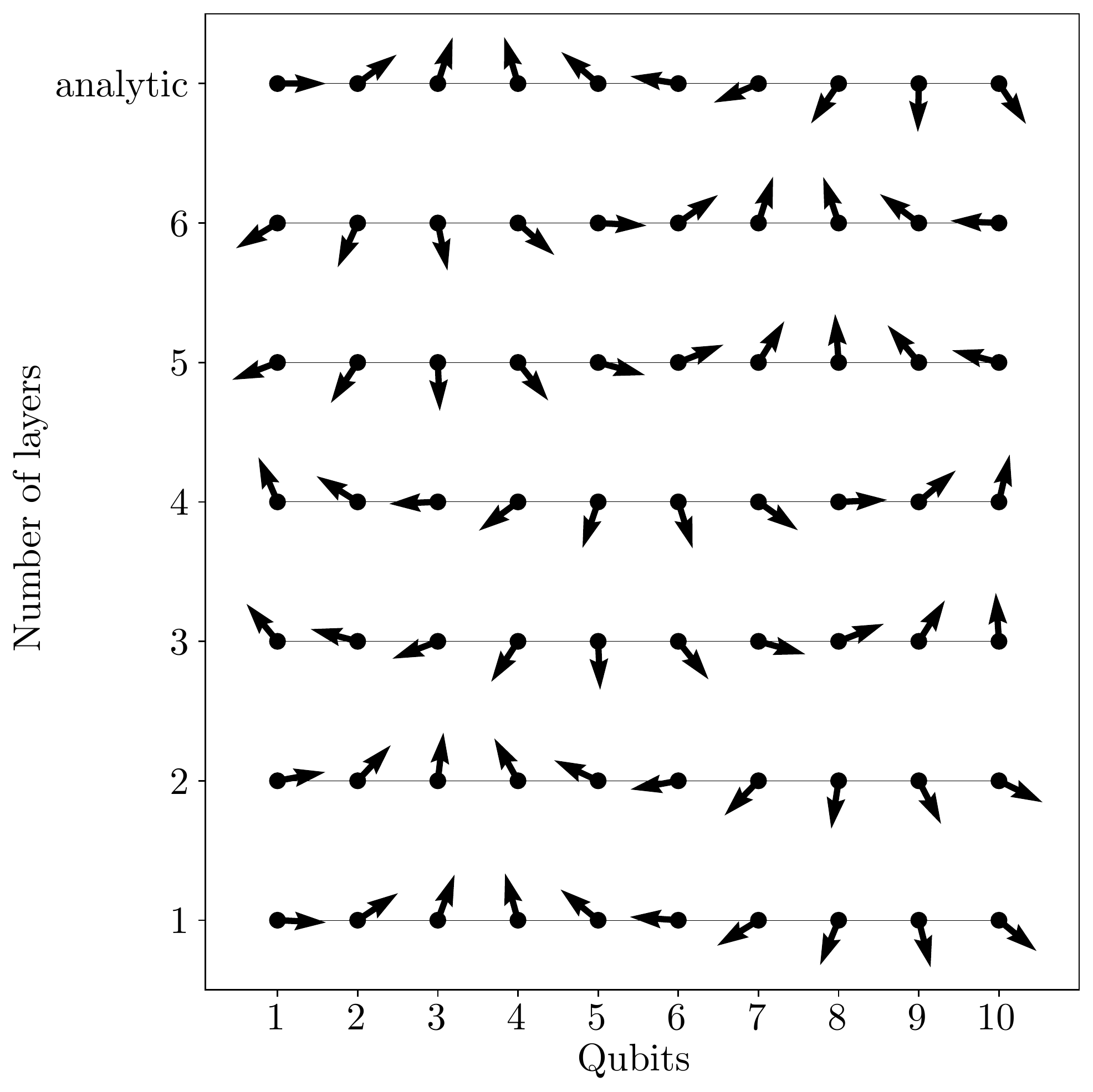}
    \caption{A magnetic texture of $N=10$ spin one-half particles that represents one period of a chiral soliton lattice depending on the number of layers in the hardware-efficient ansatz. Here, each arrow corresponds to the magnetic moment localized at a given site. Note that $z$ components of magnetization are negligible, which is in agreement with analytical findings, and the spins rotate in the $xy$ plane from site to site. Interestingly, the analytical solution as given by Eq.~\ref{eq:solution}, that minimizes the Hamiltonian (\ref{eq:inham}) in continuum limit and is marked {\it analytic}, reproduces quite well the one-layer VQE solution.}
    \label{fig:magnetic_moments}
\end{figure}
    
{\it Discussion and conclusion.} Our numerical findings shown in Fig.~\ref{fig:energy-overlap} reveal that VQE approaches the lowest energy of the Hamiltonian \eqref{eq:inham} upon increasing the number of layers with a rather tolerable accuracy. Indeed, the discrepancy between the approximated result and the exact one does not exceed 1\%. However, VQE does not perform well when approximating the corresponding eigenstate in terms of overlap with the exact solution. This can be attributed to the fact that entanglement properties of a given spin configuration are specified by the ground state exclusively, but not the Hamiltonian that VQE is designed to minimize. To justify the statement in a more rigorous way we evaluate entanglement as given by quantum concurrence and shown in Fig.~\ref{fig:concurrences}. Clearly, the VQE solution is capable of correctly reproducing the degree of entanglement between the nearest-neighboring sites, but in the meantime it does not hold for spatially separated states beyond nearest- and next-to-nearest neighbors. In contrast, a soliton solution we worked out in this Letter is highly entangled and cannot be captured within the VQE approach without a sufficiently large number of layers of an ansatz.
For illustration, we show how the magnetic texture evolves depending on the number of layers in VQE. A qubit number along the $x$ axis in Fig.~\ref{fig:magnetic_moments} selects the corresponding lattice site, so that each arrow represents a localized magnetic moment for a given site. Note that, in full agreement with analytical results, the magnetic moments are positioned in the $xy$ plane with the $z$ components being negligible. In Fig.~\ref{fig:magnetic_moments}, we show the spin configurations for up to six layers in the hardware-efficient ansatz, while the result which respects the analytical solution in continuum limit (\ref{eq:solution}) is marked {\it analytic}. Interestingly, the analytical solution $\theta=\pi/2$ and $\varphi=2\mathrm{am}(mz/\kappa,\kappa)$ fits well a one-layer VQE solution. Increasing the number of layers in VQE should in principle lead to the exact solution, which, however, cannot be achieved with shallow quantum circuits. To show that this lack of accuracy does not arise from trainability issues, in Supplemental Material, we test the VQE with the hardware-efficient ansatz on different sets of parameters for the Hamiltonian \eqref{eq:inham}. Specifically, we show that for some assignments for $D$ and $B$ the ground state is found with high precision, and the performance of VQE is dependent on the degree of entanglement between spatially separated sites of the spin chain. The latter naturally limits the application range of VQE to short-range spin configurations.

{\it Acknowledgement.} A.P. acknowledges support from the Russian Science Foundation Project No.~20-72-00044. A.K. and J.B. acknowledge support from Agreement No. 014/20, Leading Research Center on Quantum Computing.

\bibliographystyle{apsrev4-2}
\bibliography{main.bbl}

\clearpage
\onecolumngrid

\section*{SUPPLEMENTAL MATERIAL}
The Hamiltonian (1) in the main text describes a spin chain with competing pairwise interaction, namely, symmetric and asymmetric exchange couplings. The former, the ordinary direct exchange with the coupling strength $J$, is known to stabilize collinear magnetic ordering. Whereas the latter, the so-called Dzyaloshinskii-Moriya interaction with the coupling strength $D$, results in spin canting. A delicate interplay between these interactions leads to the formation of a spin-spiral magnetic structure, while upon increasing an external magnetic field $B$ this state transforms to a ferromagnet through the formation of a soliton lattice. Our numerical findings suggest that VQE with shallow hardware-efficient ansatz circuits allows rather accurate description of a less entangled spin-polarized state, which is not the case for the soliton solution. The spin-spiral magnetic structure and soliton lattice are characterized by a higher entanglement as compared to the ferromagnet state and can be captured by VQE only qualitatively.  

Here we provide the details of VQE training for a given Hamiltonian with the parameters $J$, $D$, and $B$ specifying the spin-spiral magnetic texture, soliton solution, and spin-polarized state. The results discussed in the main text are obtained using the BFGS algorithm with the number of optimization iterations limited to 50 000. Note, it was pointed out in Ref.~\cite{wierichs2020avoiding} that for a system of up to twenty spins the BFGS optimizer is a reliable approach to find a global minimum. We make use of the $l$-layered hardware-efficient ansatz to approach the soliton solution. In particular, we address the dependence of the fidelity of the VQE-solution on $l$ by varying the number of layers from one to six. The structure of the ansatz is fixed, and for each $l$ the VQE algorithm is executed independently five times with random initial parameters. Remarkably, a growing ansatz could be more effective; in this case, however, one must be aware of the fact that {\it piecewise} training (\textit{e.g.}, layer by layer) is non-efficient \cite{campos2021abrupt}.

To explore the expressibility of the hardware-efficient ansatz, extra simulations with up to nine layers have been performed.
For a set of parameters $D$ and $B$ specifying the soliton solution, the VQE converges to the state $\vert\psi\rangle$ that yields $\vert\langle\phi\vert\psi\rangle\vert^2=0.658$ with the true ground state $\vert\phi\rangle$. If, however, one directly maximizes the fidelity $\vert\langle\phi\vert\psi\rangle\vert^2$, one reaches $\vert\langle\phi\vert\psi\rangle\vert^2 = 0.9851$; we emphasize that this, up to the minus sign, is equivalent to executing the VQE for the $N$-qubit Hamiltonian $\tilde{H}=-\vert\phi\rangle\langle\phi\vert$, which has only one zero-energy excited state being $(2^N-1)$-fold degenerate. We therefore guess that with a sufficient number of layers the hardware-efficient ansatz is expressive enough for reproducing the true ground state. Interestingly, certain models, including XXZ chain, can be worked out in a closed analytical form in terms of Bethe states, {\it i.e.}, all eigenstates can be described in terms of multi-magnon states. However, the use of Bethe states as variational states appears to be limited to the case of one-magnon trial states \cite{nepomechie2020bethe}.

\begin{figure}[h!]
    \centering
    \includegraphics[width=0.45\textwidth]{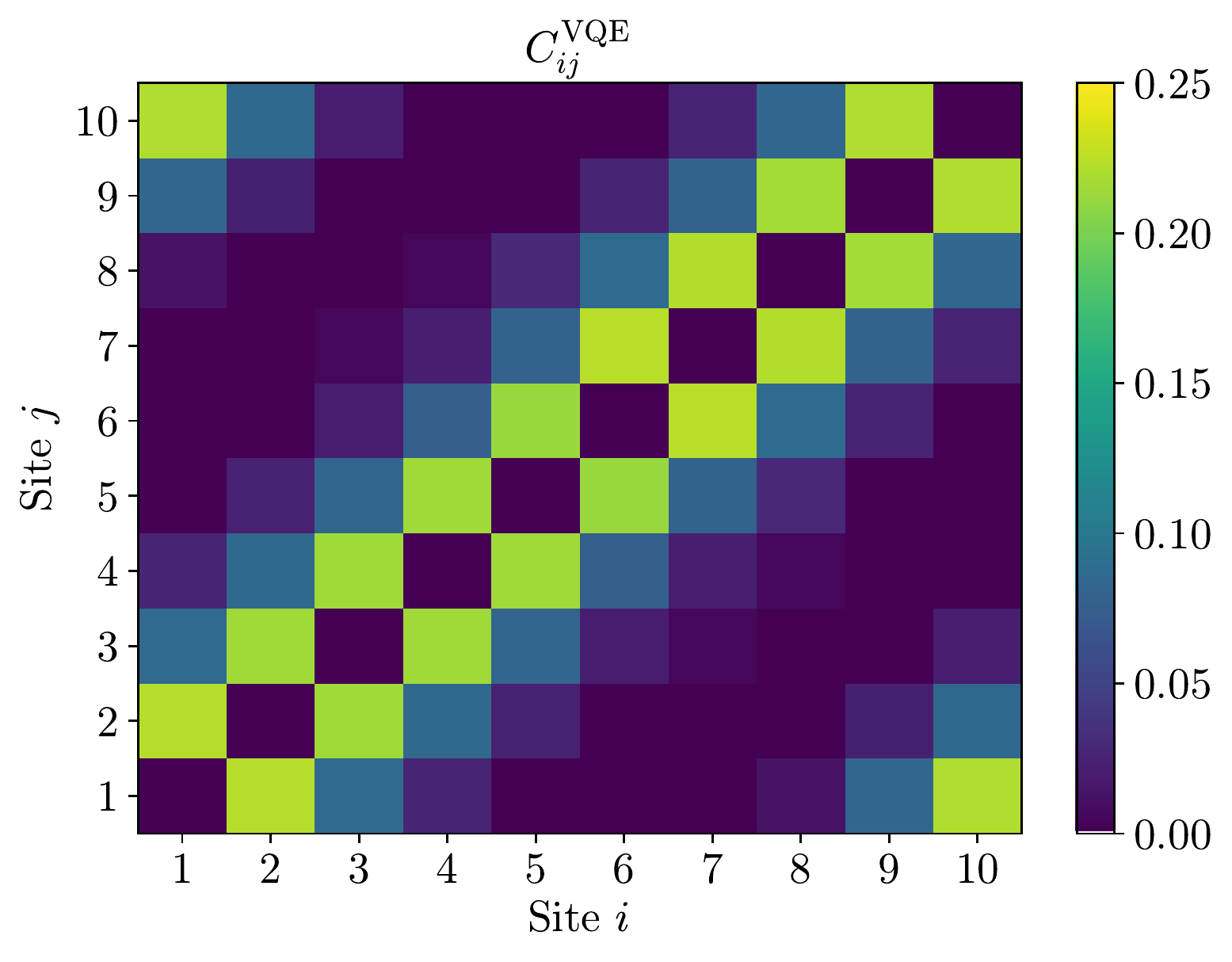}
    \includegraphics[width=0.45\textwidth]{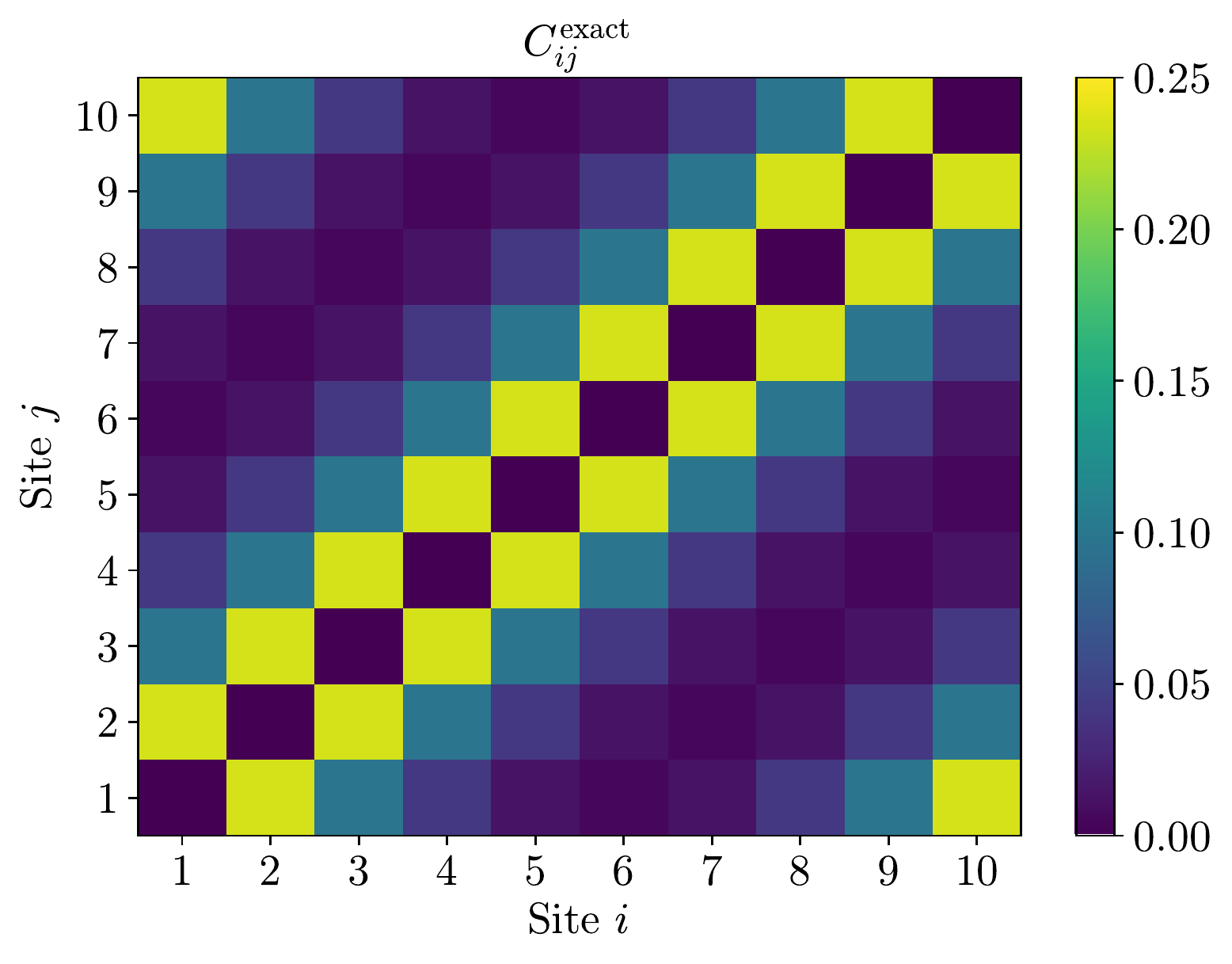}
    \caption{Concurrence based on the approximate (left) and exact (right) solutions provided $D/J=1$, $B=0$.}
    \label{fig:1D-0B}
\end{figure}

\begin{figure}
    \centering
    \includegraphics[width=0.45\textwidth]{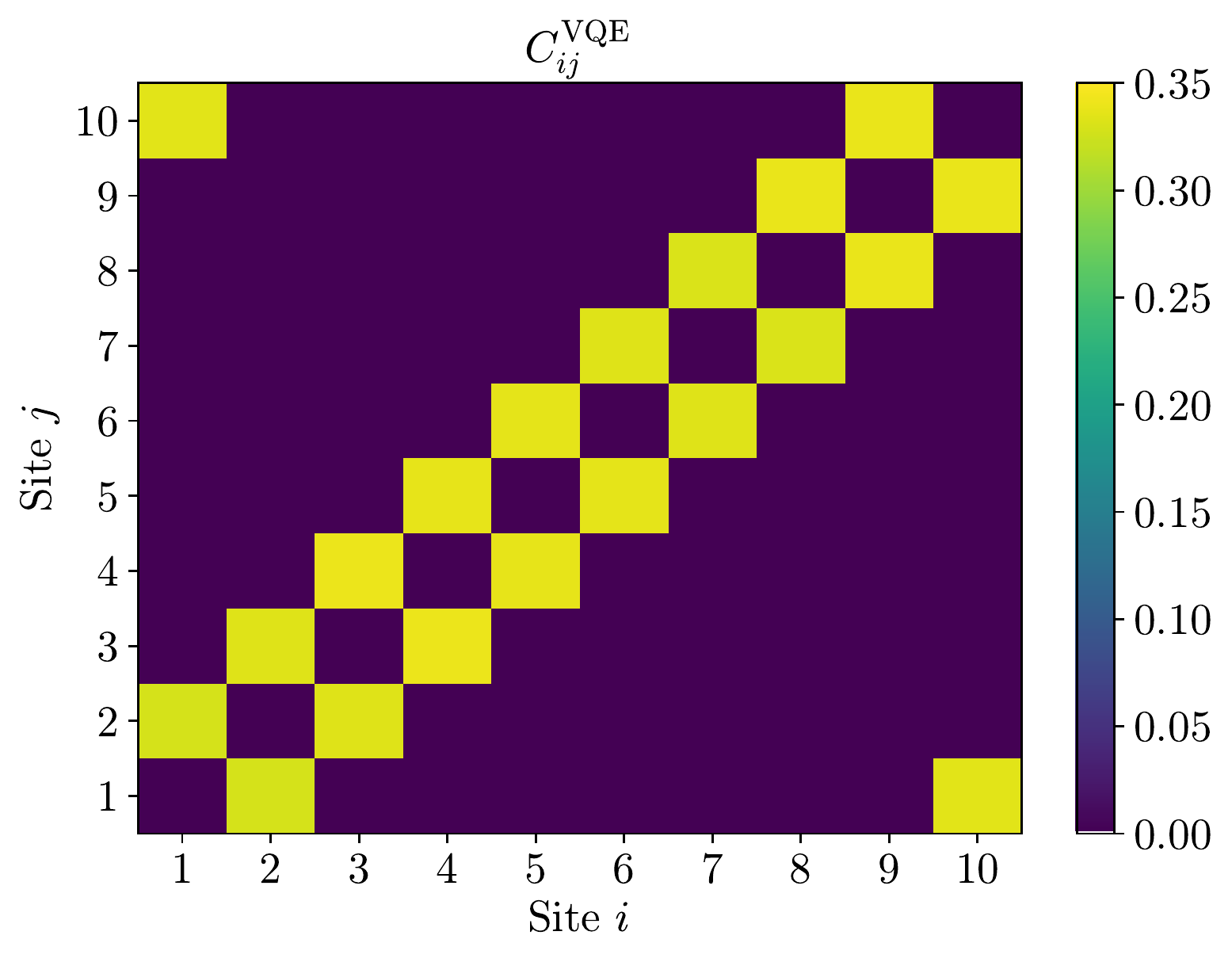}
    \includegraphics[width=0.45\textwidth]{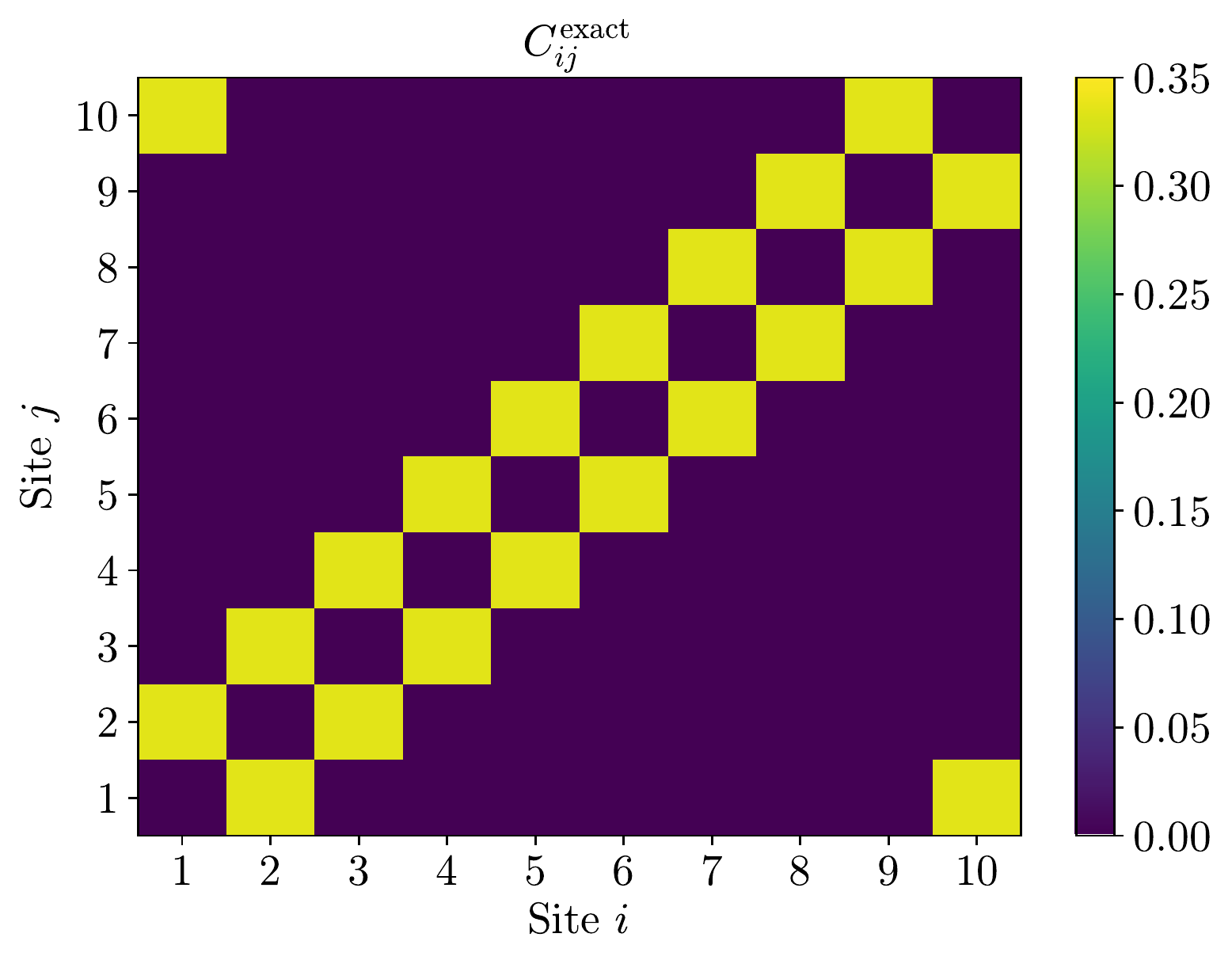}
    \caption{Concurrence based on the approximate (left) and exact (right) solutions provided $D/J=B/J=5$.}
    \label{fig:5D-5B}
\end{figure}

We focus on the parameters $D/J=0.63$ and $B/J=3.36\cdot 10^{-3}$ for $N=10$ qubits in the main text; however, for completeness of our study, we also address:

\begin{enumerate}
    \item[{(1)}] $D=B=0$. The ground state as obtained by exact diagonalization is 11-fold degenerate with at least one of them being separable. This separable state is captured by VQE with fidelity $1$ with a single-layered ansatz.
    
    \item[{(2)}] $D=0$, $B/J=1$. The ground state is unique and separable, and is obtained by VQE with fidelity 1 by means of a single-layered ansatz.
    
    \item[{(3)}] $D/J=1$, $B=0$. The solution with fidelity $0.7027$ and $\delta = 0.4017$ (which quantifies how close the VQE energy to that of exact one, see the main text) is found using an eight-layered ansatz. Concurrence of a magnetic spiral given by this solution is shown in Fig.~\ref{fig:1D-0B}.
    
    \item[{(4)}] $D/J=B/J=5$. An eight-layered ansatz results in the solution that is characterized by fidelity $0.9973$ and $\delta = 0.0358$. It turns out that only adjacent sites are entangled, as shown in Fig.~\ref{fig:5D-5B}. 
\end{enumerate}
Thus, we conclude that the performance of VQE is quite sensitive to the degree of entanglement between spatially separated parts of a system and provides better results for less entangled states.

\end{document}